# Polarization evolution dynamics of dissipative soliton fiber laser

Lei Gao[1,4], Yulong Cao[1], Stefan Wabnitz[2,3], Hongqing Ran[1], Yujia Li[1], Wei Huang[1], Ligang Huang[1], Danqi Feng[1], Xiaosheng Tang[1], and Tao Zhu[1,5]

1 Key Laboratory of Optoelectronic Technology & Systems (Ministry of Education), Chongqing University, Chongqing, 400044, China.
2 Dipartimento di Ingegneria dell'Informazione, Elettronica e Telecomunicazioni, Sapienza Università di Roma, via Eudossiana 18, 00184 Roma, Italy.
3 Novosibirsk State University, 1 Pirogova str, Novosibirsk 630090, Russia.
4 gaolei@cqu.edu.cn.

5 zhutao@cqu.edu.cn.

**Dissipative solitons emerge as stable pulse solutions of non-integrable and non-conservative nonlinear physical systems, owing to a balance of nonlinearity, dispersion, filtering, and loss/gain. At variance with conventional solitons, dissipative solitons exhibit extremely complex and striking dynamics, and their occurrence is more universal in far-from-equilibrium natural systems. Taking advantage of the techniques of the dispersive Fourier transform and time lens, the spatio-temporal transient dynamics of dissipative solitons emerging from a fiber laser have been characterized in terms of both their amplitude and phase. Yet, a full avenue of the buildup process of dissipative solitons is full of debris, for the lack of the polarization evolution information. Here, we characterize the probabilistic polarization distributions of the buildup of dissipative solitons in a net-normal dispersion fiber laser system, mode-locked by single-wall carbon nanotubes. The laser system operates from random amplified spontaneous emission into a stable dissipative soliton state as the cavity gain is progressively increased. Correspondingly, the state of polarization of each spectral wavelength convergences towards a fixed point. To reveal the invariant polarization relationship among the various wavelength components of the laser output field, the phase diagram of the ellipticity angle and the spherical orientation angle are introduced. We find that the state of polarization of each filtered wavelength in the central wavelength region of the dissipative soliton evolves linearly across its spectrum, while the states of polarization in the two fronts of the spectrum are spatially varying. Increasing cavity gain leads to spectrum broadening, meanwhile, the polarizations of the new generated frequencies extend to be scattering. Further increasing pump power results into dissipative soliton explosions, upon which a new kind of polarization optical rogue waves is identified. Those experimental results provide a deeper insight into the transient dynamics of dissipative soliton fiber lasers.**

## 1. INTRODUCTION

Solitons are spatio–temporal localized structures, associated with analytical solutions of conservative and integrable systems, where the interaction of nonlinearity and diffraction/dispersion are balanced [1]. In optics, conventional soliton laser systems possess a high degree of coherence, due to the well-defined phases among the various longitudinal laser modes. Solitons propagate with a transform-limited pulse duration, and collide with each other while maintaining their shapes and velocities. Yet, most natural systems occur to be dissipative and nonintegrable. Various localized structures, such as Akhmediev breathers, quasi-solitons, dissipative solitons, and even sporadic rogue waves, proliferate and decay with a diverse range of spatial-temporal periodic and localized structures [2-4].

Dissipative solitons (DSs) are high coherent solutions of nonlinear wave equations, and arise from a balance between nonlinearity, dispersion, filtering, and loss/gain [1-4]. At variance with conventional soliton in integrable fiber laser systems with anomalous dispersion, DSs in dissipative fiber laser systems operating in normal dispersion exhibit extremely complex and striking dynamics. For example, their much broader (when compared with solitons) pulse durations and linear phase chirping make them ideal for high energy ultrafast fiber lasers and phase-dependent optical processing. Moreover, their flat and broad optical spectrum provides an excellent platform for high coherent laser sources. Extensive theoretical analyses of DSs in mode-locked fiber laser systems have been performed [5], while direct experiments on their buildup dynamics are often restricted by the relative slow detecting techniques. Based on the well-developed dispersive Fourier transform (DFT) method, non-repetitive spectra of optical signals are transformed into temporal data under the far-field approximation. As a result, the spectral properties of DSs, soliton explosions, soliton molecules, modulation instability, and supercontinuum have been characterized [6-14]. Similarly, the time–lens method, which utilizes the temporal analogue of a spatial lens, has been exploited for the characterization of real-time pulse temporal properties of transient DS [15,16]. However, the consistence of amplitude and phase measurements in spatio-temporal transient dynamics relies on the assumption of a strict correspondence between temporal and spectral domains. The validity of such assumption is suspicious in transient optical dynamics, where the optical phases may be loosely locked, or even irrelevant. This is especially important in the proliferating period of DSs, where, for example, the state of polarization (SOP) of laser longitudinal modes at any point in time may be a random variable [17-22]. This lack of completeness makes the full avenue of the buildup dynamics of DS full of debris.

In this letter, we experimentally investigate the probabilistic nature of the SOP distribution in the buildup dynamics of DSs. The DS proliferates from random noise in a fiber ring laser cavity with net normal dispersion, which is mode-locked by a saturable absorber (SA) made from single-wall carbon nanotubes. We characterize the probabilistic

SOP distribution by introducing the phase diagram of the ellipticity angle χ and of the spherical orientation angle ψ. We reveal an inversely exponential distribution of the SOP with increasing cavity gain, and investigate the spatially varying structure of the various spectral components of a stable DS. A new kind of polarization optical rogue waves is also identified, when the high coherence of the DS is deteriorated by imposing a progressively higher cavity gain.

## 2. MAIN PRINCIPLE AND SETUP

Figure 1 schematically depicts the ring fiber cavity containing a 15 m erbium-doped fiber (EDF, Nufern, EDFC-980-HP), forward pumped by a 976 nm continuous wave laser through a wavelength division multiplexer (WDM), a polarization independent optical isolator (ISO), a polarization controller (PC), 2.7 m of single mode fiber (SMF), and an optical coupler (OC) with a 10% output port. The dispersion parameters of the EDF and SMF are -12.2 and 18 ps/nm/km, respectively. Therefore, the total 17.8 m ring cavity possesses a net normal dispersion of 0.171 $ps^2$. The saturable absorbing film is made by mixing 10 wt% aqueous polyvinyl alcohol and 0.5 mg/mL single-wall-carbon-nanotubes solution at a volume ratio of 1:2, then dehydrated in vacuum [23]. Its nonlinear optical response is measured by means of a home-made balanced two-detector, as described in [24]. The modulation depth and the saturation power are 3.5% and 30.73 $MW/cm^2$, respectively [23]. This low saturation power permits our laser system to operate in a DS regime with relatively low pump powers. At the same time, the DS become unstable at relatively high pump powers, owing to the bleaching effect of the SA.

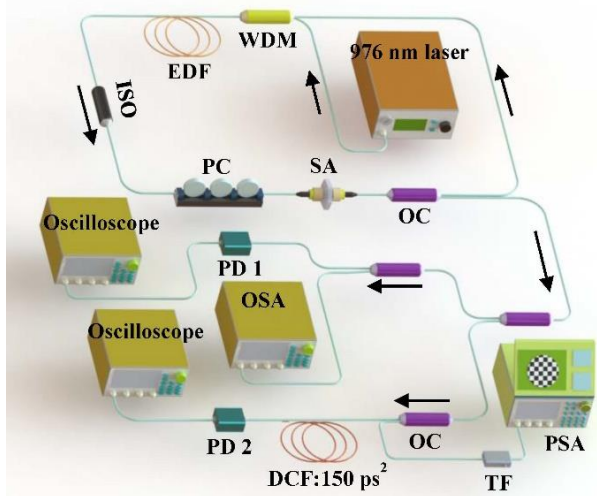

**Fig. 1.** Schematic of the ring fiber laser cavity and measurement methods.

The laser output is characterized by an optical spectrum analyzer (OSA, Yokogawa, AQ6370), an autocorrelator (APE, Pulse check), an oscilloscope (Lecroy, SDA 8600A), and a frequency analyzer (Agilent, PSA E4447A), together with a photodetector (PD1) with a bandwidth of 350 MHz. The SOP of the DS is measured by a high-speed polarization state analyzer (PSA, General Photonics, PSGA-101-A) after filtering by a tunable optical filter (TF, Santec, OTF-320) with a bandwidth of 0.2 nm. Meanwhile, we also detect the laser single-shot spectra by means of a home-made DFT, where periodic signals are stretched by a 500 m dispersion compensation fiber (DCF) with the dispersion of 150 $ps^2$ for the frequency-to-time transformation, and subsequently fed to a 50 GHz PD2 connected to a real time oscilloscope (Tektronix, DPO 71254) with the bandwidth of 12.5 GHz.

## 3. RESULTS AND DISCUSSION

For the configuration in Fig.1, stable DS can be observed for pump power larger than 55 mW with easiness. The typical output parameters for a pump power of 55 mW are depicted in Fig. 2(c), where the rectangle-shaped optical spectrum with a full width at half maximum (FWHM) of 13.6 nm is shown. The autocorrelation trace in Fig.2(e) exhibits a FWHM duration of 30.5 ps, when fitted by a Gaussian function. The signal to noise ratio at the fundamental frequency of 11.48 MHz exceeds 70 dB. We also plot the optical spectra under various pump powers in Figs. 2(a)-(d). For pump power smaller than 55 mW, the laser system operates from random amplified spontaneous emission into a quasi-DS regime, until it inters into a stable DS regime. Further increasing the pump power broadens the optical spectrum, until a too large pump-induced gain deteriorates the mode-locking state. This deterioration can be deduced directly from the noisy new-emerged frequencies in the two fronts of DS optical spectrum in Fig. 2(d), which originate from stochastic longitudinal laser modes mixing via the nonlinear optical response of the fabricated SA with a low saturation power.

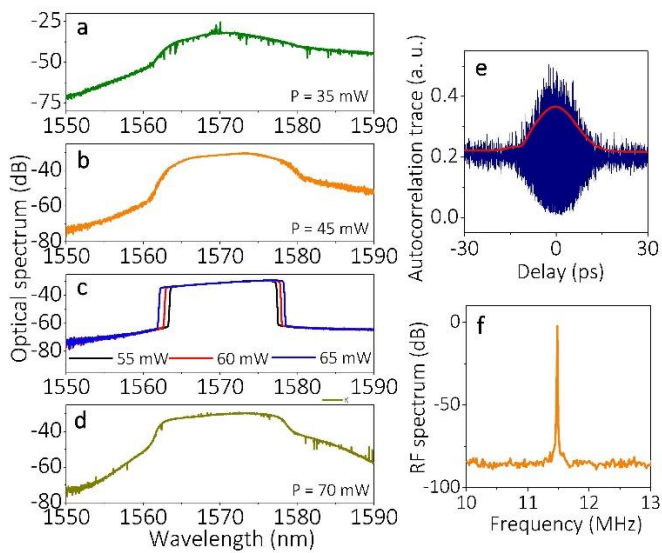

**Fig. 2.** Typical outputs for DS. (a)-(d) Averaged optical spectra for various pump powers. (e) Autocorrelation trace of pulses for pump power at 55 mW, which is fitted by Gauss function. (f) Radio frequency spectrum at the fundamental frequency of 11.48 MHz.

In order to further explore the intrinsic transient dynamics of the laser cavity, in Fig.3 we display consecutive single-shot optical spectra. Real-time optical spectra corresponding to those in Figs. 2 (a), (b), (d) exhibit fine structures, which are invisible in the averaged optical spectrum. For example, Fig. 3 (a) shows that a broad optical spectrum is formed for a pump power of 35 mW. In fact, the laser system includes both a pulsed component, together with coupling to a continuous wave (CW) component. The DFT technique can only display the dynamics of the pulsed component. As we can see, the optical spectrum grows larger by drawing energy from the CW component, until it decays swiftly due to the insufficient cavity gain, similarly to the optical puff which is gestated in a partially mode-locked laser system [11,13]. Interestingly, Fig. 3 (a) shows the generation of new frequencies in the short wavelength side of the laser spectrum, owing to wave mixing among the bright spectral stripes, while a giant spectral feature appears in the central wavelength region. Just following the formation of a spectral peak, the whole broad spectrum collapses swiftly, until laser energy returns back to the CW component, resulting into a near zero-value field mapping when detected by the DFT.

In contrast with the previous behavior, Fig. 3(b) shows that, as the pump power grows to 45 mW, the optical spectrum gradually broadens, until two giant peaks appear in the two fronts of the spectrum. Those peaks draw their energy from the CW component, and they progressively red-shift and blue-shift as the number of circulations grows larger, respectively. Meantime, multiple mode mixing occurs between the two peaks, as it can be also seen in Fig. 3 (e). After reaching a peak value, the optical spectrum decays under further cavity circulations, albeit with a slower rate when compared with that of Fig. 3 (a). The energy of the pulsed component flows back into the CW component.

Until now, cavity gain is not sufficient yet to support stable DS, so that irregular clusters proliferate and decay at various rates in a memoryless manner, which only depends on the pump power level. Yet, our results show that great difference exists between a partially mode-locked laser and DS laser [13]. For example, we do not find the presence of critical behavior in the buildup of DSs. Typically, this behavior means that the proliferation time and decay time are the same, as it has been observed for the optical puff associated with the onset of optical turbulence [13]. The reason is that underlying physics of the two types of lasers are different: the evolution of the optical puff in a partially mode-locked fiber laser system is only associated with the presence of multiple-mode-mixing [11,13]. Whereas the formation of an optical cluster in a dissipative fiber laser system is more complex, involving both nonlinearity, dispersion, filtering, and loss/gain. In its decaying regions, energy does not gradually couple into the central part of the spectrum as it occurs for the optical puff, but rather it collapses swiftly into the CW component.

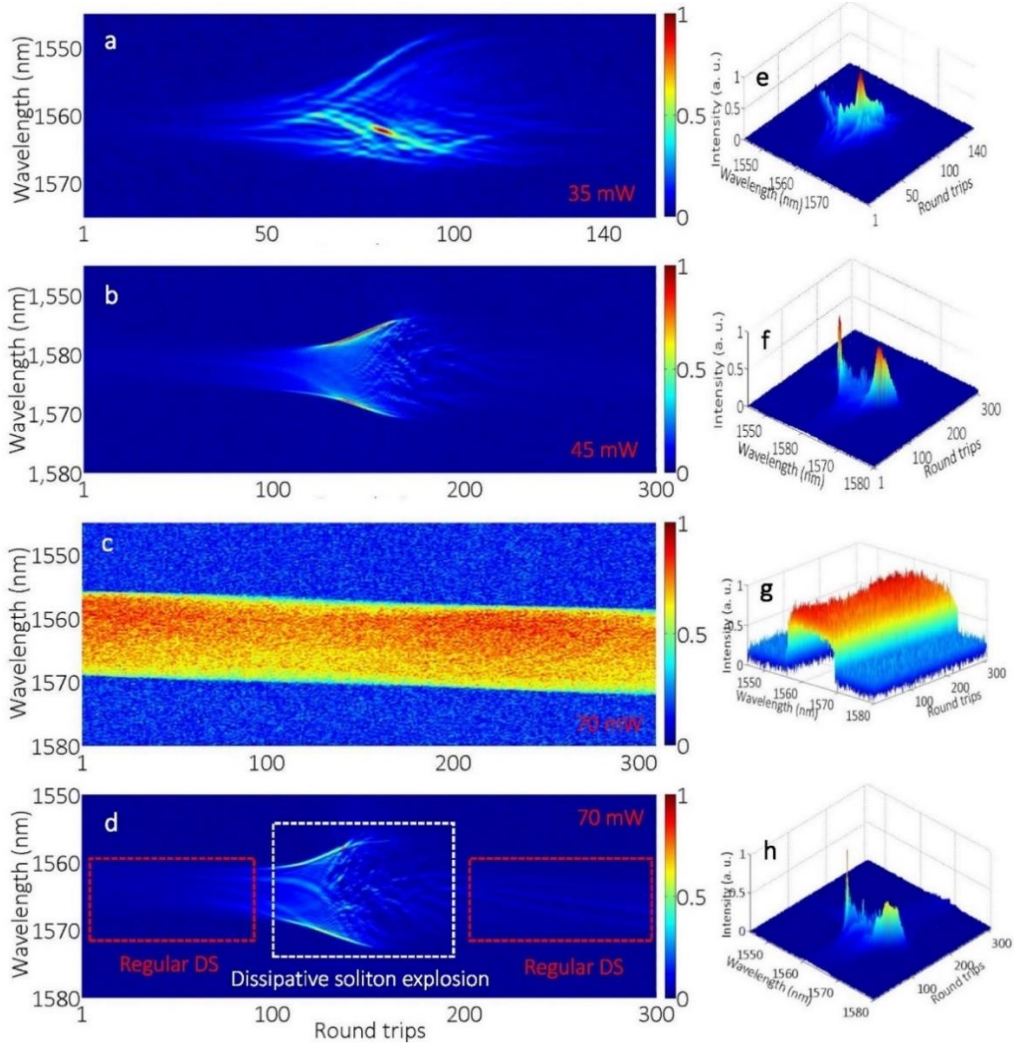

**Fig. 3.** Conservative single-shot spectra in the build-up of DS at various pump powers. (a)-(d), Pump power at 35 mW, 45 mW, and 70 mW, respectively. The corresponding 3D plots in (e)-(h) indicate giant peaks with rogue intensities in the spectral domain. Due to limited splicing error of long time series into pieces, a slight tilt is shown in the single-shot spectra. All intensities are normalized with respect to the maximum value in each detection. The varying bright stripes intersect within successive roundtrips, indicating intense multimode mixing.

For pump power between 55 mW and 65 mW, the laser operates in a stable DS regime. Yet, its average intensity during single-shot spectrum detection is much smaller than those of unstable pulses as in Figs. 3(a), (b) and (d). Due to the large formation time, and the limited storage length of the real-time oscilloscope, we could not detect the process of formation of a stable DS. Yet, this information has been carefully studied via single-shot detection based on DFT [4,8]. Instead, our aim here it to study the deterioration of the DS state, as it is induced by excessive cavity gain. This phenomenon has been frequently encountered when the saturation power of the SA is relatively low, as it occurs in our case with single-wall-carbon-nanotubes. As shown in Figs. 3 (c) and (g) for a pump power of 70 mW, regular DS with neatly rectangle-shaped optical spectra are frequently detected, just as it occurs for pump power between 55 mW and 65 mW. Yet, much broader optical spectra persisting near the square spectrum may also be occasionally encountered, exhibiting two extremely high peaks. One such example is shown in Figs. 3 (d) and (h). After the spectral collapse, the DS returns to its neat square shape with a much lower intensity. In this case, the whole transient evolution process takes place within 50 roundtrips. Limited by the low dynamic range of the high speed PD2, the single-shot spectrum of the stable DS is too low when compared with the extremely high intensities of the transient stage. This phenomenon is referred to as *soliton explosion*, whereby the balance of nonlinearity, dispersion, filtering, and loss/gain for the DS is perturbed by the surplus cavity gain at large pump powers [25]. Part of the DS energy dissipates into CWs via the explosion, and the DS maintains its property of a high coherence pulse. Based on thousands times of detections, we find that the probability for these giant peaks increases with cavity gain, and the transient evolution will take a longer time to return into the DS region.

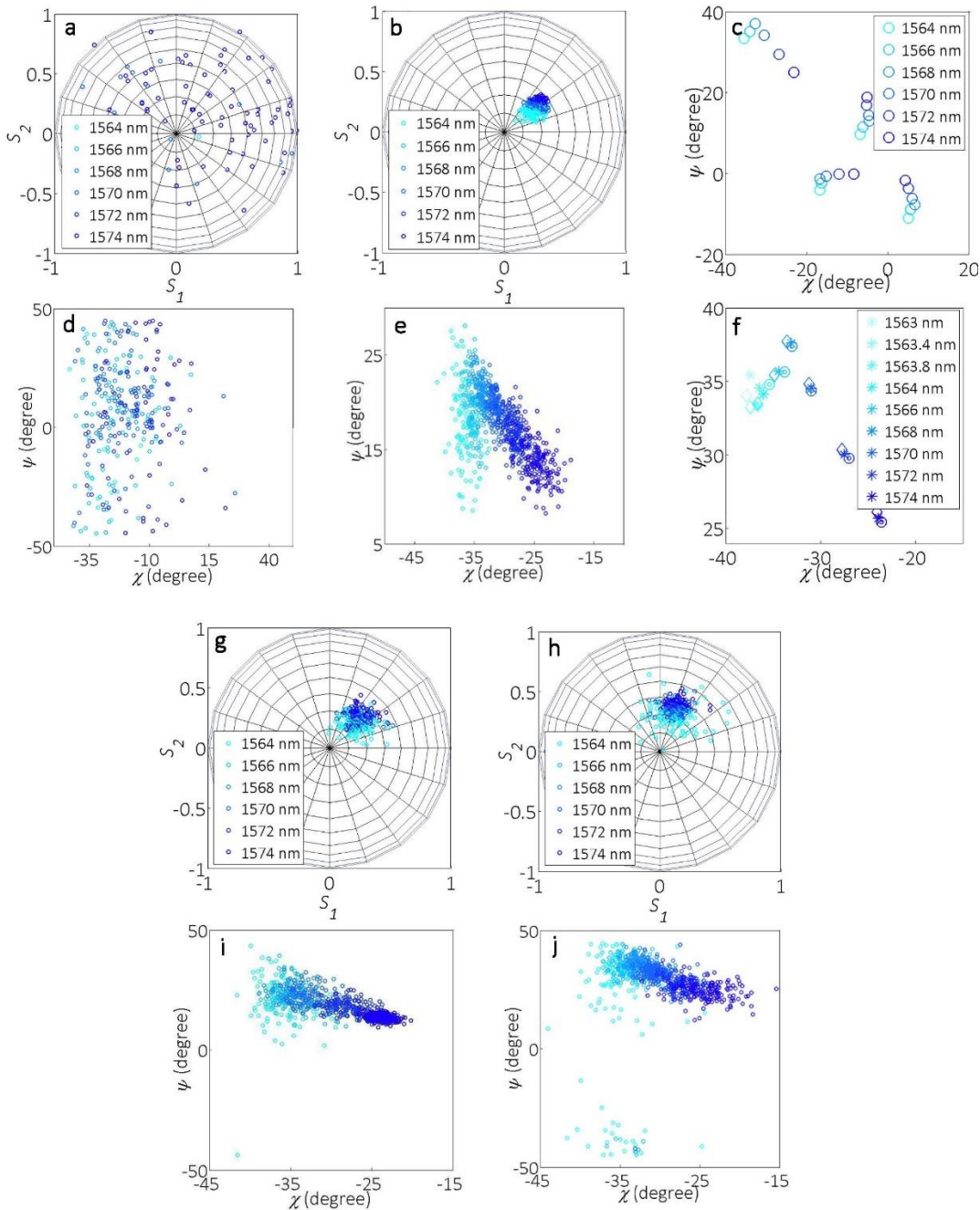

**Fig. 4.** States of polarization for various filtered wavelengths and different pump powers. (a), (b), (g), (h) Normalized Stokes parameters represented on the Poincaré sphere for pump power of 35 mW, 45 mW, 67.5 mW, and 70 mW, respectively. The corresponding phase diagrams based on the ellipticity angle $\chi$ and spherical orientation angle $\psi$ are shown in (d), (e), (i), (j). (c) Phase diagram for pump power at 55 mW, where the absolute SOP entering the PSA is changed arbitrarily through biasing the fiber between output OC and the PSA. (f) Phase diagram for a stable DS, where the pump power is 55 mW (circled dots), 60 mW (rectangle dots), 65 mW (star dots), respectively. The SOPs on the edges of the spectrum exhibit fluctuations.

Our experimental data arise a question: why cannot the DS further broaden its spectrum indefinitely, as the pump power is progressively raised, as it occurs in the case of a DS resonance [26]? To ascertain the response to this question, we need to reconsider how new frequencies are generated in the process leading to a DS, from stochastic amplified spontaneous emission (ASE) noise, in the process leading to the well-known linear phase relationship among the frequency components of a DS. Although the DS single-shot spectra have been previously reported, the characteristics of the exact SOP distribution of the longitudinal modes of the laser may provide an additional insight for reconstructing the buildup process of a DS. This information is especially important in a primary build-up stage, where the DFT fails due to random phase relationship among the laser modes.

Therefore, we measured the SOPs of different filtered wavelengths from the dissipative fiber laser cavity, for different levels of the pump power. Figure 4 exhibits the resulting SOP distributions for the cases in Fig. 2. It is clear from the distribution of points on the Poincaré sphere that the corresponding SOPs for each wavelength are evolving from a random cloud into a fixed point as the pump power grows larger. This trend is more apparent in a phase diagram for the ellipticity angle χ versus the spherical orientation angle ψ, which are calculated as follows

$$\begin{cases} \chi = \frac{1}{2}\arctan(\frac{s_3}{\sqrt{s_1^2 + s_2^2}}) \\ \psi = \frac{1}{2}\arctan(\frac{s_2}{s_1}) \end{cases} \quad (1)$$

Here, $s_1, s_2, s_3$ are the stokes parameters of the laser SOP. At variance with the case of partial mode-locking [11], neither a bifurcation nor total turbulence can be identified here. As shown in Fig. 4 (d), the positions of the SOPs follow a random distribution for 35 mW of pump power, where ASE dominates. Yet, Fig. 4 (e) shows that for the pump power of 45 mW all SOPs are located in a well-defined region of the phase diagram for χ and ψ, although they still behave erratically.

Further increasing the pump power leads to stable DS operation. As shown in Fig. 4 (c), a well-defined, rectangle-shaped optical spectrum results into a fixed SOP distribution with a folding line. The distribution in the phase diagram of χ and ψ remains invariant, even if the birefringence induced by the fiber between the output OC and the polarization state analyzer is changed. We observe a quasi-V shaped SOP distribution in Fig. 4, even in cases where the DS is not stable because of insufficient cavity gain. The SOP phase diagram for wavelengths in the central region of the DS shows linearly spaced values. Whereas wavelengths in the blue side of the DS spectrum show a similar behavior, but in a different direction. The reason is the wavelength-dependence of the filtering film in the TF. Limited by the manufacturing process of the tunable filter, we can only tune the wavelength continuously from 1563 nm to 1574 nm, beyond which a sudden abrupt switching of the SOP for longer wavelengths is added by the TF. Yet, we observe a similar distribution in the phase diagram for wavelength larger than 1575 nm. In other words, both long and short wavelength sides of the DS spectrum exhibit spatially varying SOP distributions.

In general, larger cavity gains will broaden the optical spectrum of a stable DS, by generating additional frequencies on both the long and the short wavelength side of its spectrum. This spectral broadening can be verified in Fig. 2 (c), and its corresponding phase diagrams as depicted in Fig. 4 (f). We find that new wavelengths on both the long and short wavelength sides of the DS spectrum exhibit irregular SOP values, that depart from the distribution of the center of the DS spectrum. Moreover, also SOPs corresponding to the central region of the DS spectrum are shifted slightly with pump power.

For example, when the pump power is 65 mW, new frequencies on the edges of the spectrum become unstable, even though the DS remains stable as a whole. Those experimental results point to a change of the energy redistribution of among lasing wavelengths, as the cavity gain is changed. Nevertheless, such a phenomenon has hardly been taken into account when building theoretical models for dissipative mode-locked laser systems. Besides, the occurrence of a spatially varying SOPs on the two edges of the spectrum has not been fully considered when using DS for practical applications, such as, pulse compression, amplification, pump-probing, and supercontinuum generation.

A more general problem that is frequently encountered is the DS deterioration when the cavity gain becomes large enough. In our case, this means for pump power beyond 70 mW: the corresponding optical spectrum is shown in Fig. 2 (d). Its corresponding SOP information on the Poincaré sphere and in the phase diagram are depicted in Figs. 4(g)-(j). During the soliton explosion, it is clear that more frequencies with freak or rogue positions in the phase diagram are observed when the DS is deteriorated by larger cavity gain. Besides, the probability for generating such freak frequencies seems to increase as the gain grows larger. This trend is obvious when comparing the single-shot spectra in Figs. 3 (c) and 3 (d), where new frequencies on the two edges of the soliton spectrum are generated with unexpected high intensities. During the dissipative soliton explosion, the system cannot be self-sustaining. Also, those new frequencies are fundamentally different from those generated when the cavity gain is insufficient. For example, quasi-DSs are frequently encountered for a pump power at 45 mW, while no freak SOP distribution has been observed for the whole lasing spectrum.

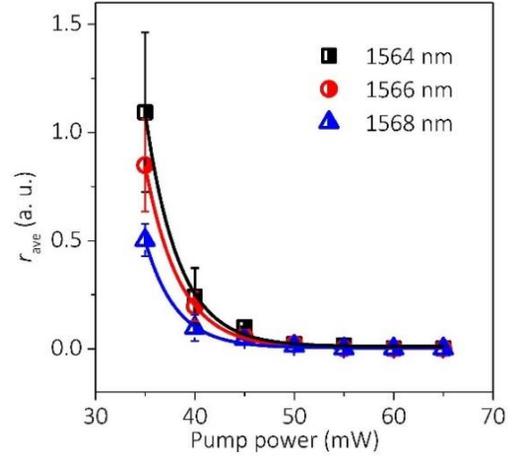

**Fig. 5.** The average relative distances of the laser wavelengths SOPs for different pump powers. Exponential convergence towards a fixed point is observed. Error bars represent absolute errors based on experimental data, which are relatively high for lower pump powers.

To quantitatively characterize the SOP distributions in the buildup of a DS, we provide a dimensionless parameter, $r$, describing the relative distance between two arbitrary points of the distribution, representing two different Stokes vectors:

$$r = \sum_{\substack{m,n=1 \\ m \neq n}}^{N} \left| \hat{S}_m - \hat{S}_n \right| = \sum_{\substack{m,n=1 \\ m \neq n}}^{N} \sqrt{(s_{m1} - s_{n1})^2 + (s_{m2} - s_{n2})^2 + (s_{m3} - s_{n3})^2} \quad (2)$$

We calculated the average value of $r$ for different cavity gains. In Figure 5 we plot $r_{ave}$ for various filtered wavelengths in the blue-shifted portion of the DS spectrum (curves in the red-shifted portion are not shown, as they are symmetric to the blue shifted ones). As can be seen, for each wavelength the relative distance $r_{ave}$ convergences exponentially with pump power into a fixed point. Wavelengths away from the central region of the DS spectrum exhibit reduced convergence rates.

Next, let us consider the probability distribution of $r$ when the DS is deteriorated at high pump powers. As seen in Fig. 6, the quasi-Gaussian shape distribution for the wavelength at 1568 nm (the center region of the DS spectrum) originates from the DS, although may be disturbed by the dissipative soliton explosions. Howbeit, for wavelengths far away from the DS spectral center, a trend develops towards L-shaped probability distributions, which characterize the emergence of extreme events in the polarization domain, rather than in time or frequency domains [19,27-31]. The irregular polarization state of a deteriorated DS is associated with the emergence of a new type of optical rogue waves in the polarization dimension, namely optical polarization rogue waves [13]. From the point of view of their measurement in the phase histogram in Figs. 4 (i) and (j), such rogue events appear with both unexpected positions and relatively large probabilities of occurrence [19]. From a statistical point of view, the occurrence of polarization rogue waves can also be testified by emergence of a *heavy tail* in the measured histogram. We present in Fig. 6 the significant wave height (SWH), which is defined as the mean amplitude of the highest third of the waves. As shown, 1.8 % of the events have a value that is larger than the twice of the SWH for the wavelength at 1568 nm. Whereas rogue events represent about 3.8 % of the events for the wavelength at 1564nm. This reasoning leads to anticipate that the probability of rogue events could be even larger as the cavity gain (or pump level) is further increased.

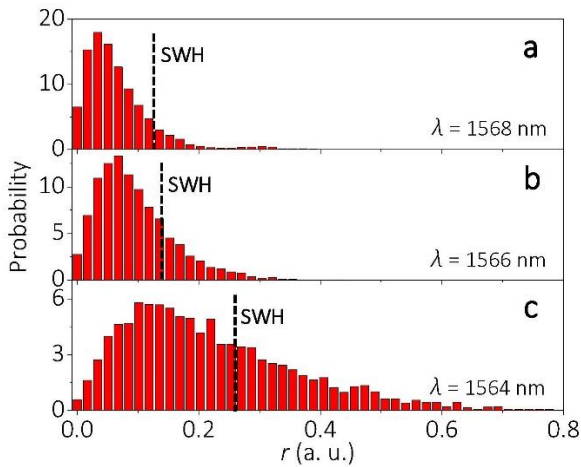

**Fig. 6.** Histograms for the relative distance between points on the Poincaré spheres for a 70 mW pump power, when a soliton explosion deteriorates the high coherence of a DS. Large deviations from a Gaussian distribution indicate the presence of an optical polarization rogue waves, which can also be certified by the SWH.

## 4. CONCLUSION

We experimentally investigated the dynamics of the polarization evolution of DS fiber lasers, mode-locked by a single-wall carbon nanotube. When the cavity gain is increased, the laser system operates from random ASE to a stable DS state. Correspondingly, the SOPs of the laser modes converge into a fixed point, with an exponential convergence rate as the pump power is increased. By introducing the phase diagram of the ellipticity and spherical orientation angles, we find that the SOP of each wavelength in the central region of the DS is linearly spaced, while the SOP in the two edges of the spectrum are spatially varying. Larger gains lead to optical spectrum broadening, and the new generated SOPs extend to scattering. Further increases of the pump power results into DS deterioration, and lead to the observation of soliton explosions. We identified the occurrence of a new type of polarization optical rogue wave, by observing a strong heavy tail in the measured SOP histogram through the SWH method. These experimental results provide a deeper insight into the transient dynamics of DS fiber lasers, which can be of great interest for theory modelling, and for technological applications of fiber lasers, optical amplification, pump-probe spectroscopy techniques, supercontinuum sources, etc. The observation of soliton explosions and polarization rogue waves in DS lasers also provide a new insight into the still hotly debated topic of the mechanisms for rogue wave generation.

**Acknowledgment**. This work was supported by the Natural Science Foundation of China (61635004, 61405023), the National Postdoctoral Program for Innovative Talents (BX201600200), the Postdoctoral Science Foundation of China (2017M610589), the Postdoctoral Science Foundation of Chongqing (Xm2017047), the Science Foundation of Chongqing (CSTC2017JCYJA0651), the Fundamental Research Funds for the Central Universities (106112017CDJXY120004), the National Science Fund for Distinguished Young Scholars (61825501). Stefan Wabnitz acknowledges support by the Italian Ministry of University and Research (MIUR) (2015KEZNYM), the European Union's Horizon 2020 research, the innovation program under the Marie Skłodowska-Curie grant agreement (691051), and the Russian Ministry of Science and Education (14.Y26.31.0017).